# Mean-free path effects in magnetoresistance of ferromagnetic nanocontacts

A.N. Useinov[1],, L.R. Tagirov[1], R.G. Deminov[1] Yun Zhou[2] and G. Pan[2]

[1] Kazan State University, Kremlevskaya str. 18, Kazan 420008, Russian Federation,
[2] CRIST, University of Plymouth, Plymouth, Devon PL4 8AA, United Kingdom

**Abstract.** We investigated the mean-free path effects on the magnetoresistance of ferromagnetic nanocontacts. For most combinations of parameters the magnetoresistance monotonously decreases with increasing the contact cross-section. However, for a certain choice of parameters the calculations show non-monotonous behavior of the magnetoresistance in the region in which the diameter of the contact becomes comparable with the mean-free path of electrons. We attribute this effect to different conduction regimes in the vicinity of the nanocontact: ballistic for electrons of one spin projection, and simultaneously diffusive for the other. Furthermore, at certain combinations of spin asymmetries of the bulk mean-free paths in a heterocontact, the magnetoresistance can be almost constant, or may even grow as the contact diameter increases. Thus, our calculations suggest a way to search for combinations of material parameters, for which high magnetoresistances can be achieved not only at the nanometric size of the contact, but also at much larger cross-sections of nanocontacts which can be easier for fabrication with current technologies. The trial calculations of the magnetoresistance with material parameters close to those for the Mumetal-Ni heterocontacts agree satisfactorily with the available experimental data.

**PACS.** 72.25.Ba Spin polarized transport in metals; -73.63.Rt Nanoscale contacts; -75.47.De Giant magnetoresistance

## 1. Introduction

Magnetic point contacts showing an extraordinary high magnetoresistance (MR) [1-7] offer a feasible avenue to a new generation of nanosize read heads for computer hard disks [8]. Theoretical studies of the MR phenomenon in magnetic nanocontacts [2,9-14] were focused mainly on the contact size and conduction-band spin-polarization effects on the MR. Other important material parameters such as spin-dependent mean-free paths (MFP) of conduction electrons in contacting ferromagnetic metals had not been taken into account. At the same time, it is well known for cobalt and permalloy that the bulk spin-up and spin-down conduction electron MFP may differ one from the other by up to 7 times (see, for example, Refs. [15,16]). Then, at a certain contact size, one conductance spin channel can be in the ballistic conductance regime, while the second spin channel still remains in the diffusive conduction regime. The situation is complicated even more if a nanocontact is realized between two non-identical ferromagnets (ferromagnetic heterocontacts) [10,17,18]. The aim of this paper is to study the mean-free path effects on the MR of magnetic nanocontacts.

## 2. Magnetoresistance in ferromagnetic nanocontacts - mean-free path effects

In a previous paper [19], we developed the quasiclassical theory of electric transport through magnetic nanocontacts. The theory is most general in considering the physical parameters of contacting ferromagnetic metals: the ferromagnets can be either identical or different, the Fermi momenta of conduction electrons spin subbands, as well as spin-dependent MFP of both ferromagnets can be arbitrary. In the paper [19] we studied in detail the influence of spin polarization and the mutual disposition of conduction bands for the case of ferromagnetic heterocontacts. To investigate effects of spin-dependent MFP we retrieve at first basic formulas for the conductance of a nanoscopic heterocontact obtained in Ref. [19]. The conductance for one of the two spin channels of conduction through the nanocontacts reads:

$$\sigma_{LR\alpha}^{P(AP)} = \frac{e^2 \left(p_F^L\right)^2 \pi a^2}{\pi^2} \int_0^\infty d\kappa \frac{J_1^2(\kappa a/l_{L\uparrow})}{\kappa} F(\kappa), \qquad (1)$$

where

$$F(\kappa) = \langle x_L D \rangle_{\theta_L} - \left(G_1 \langle x_L I_L \rangle_{\theta_L} + G_2 \langle x_L I_R \rangle_{\theta_L}\right), \qquad (2)$$

$$G_1 = \left\{\langle D \rangle_{\theta_L} [2(1-\lambda_R) + \tilde{\lambda}_2] - \langle D \rangle_{\theta_R} \tilde{\lambda}_4\right\} \Delta^{-1}(\kappa), \qquad (3)$$

$$G_2 = \left\{\langle D \rangle_{\theta_R} [2(1-\lambda_L) + \tilde{\lambda}_1] - \langle D \rangle_{\theta_L} \tilde{\lambda}_3\right\} \Delta^{-1}(\kappa), \qquad (4)$$

$$\Delta(\kappa) = 4(1-\lambda_L)(1-\lambda_R) \\ + 2\left[\tilde{\lambda}_1(1-\lambda_R) + \tilde{\lambda}_2(1-\lambda_L)\right] - \tilde{\lambda}_3\tilde{\lambda}_4 + \tilde{\lambda}_1\tilde{\lambda}_2. \qquad (5)$$

In the above formulas the subscript $L$ or $R$ refers to the left- or right-hand side of the contact, $\kappa = k l_{L\uparrow}$ and $k$ are the renormalized wavenumber and the wavenumber of an electron, respectively. They lie in a plane perpendicular to the general direction of the current flow. The other notations are as follows:

$$\langle x_L I_L \rangle_{\theta_L} = \int_{x_c}^{1} dx_L \frac{x_L D(x_L)}{\sqrt{1+\left(\kappa R_\alpha^L\right)^2 \left(1-x_L^2\right)}}, \qquad (6)$$

$$\langle x_L I_R \rangle_{\theta_L} = \int_{x_c}^{1} dx_L \frac{x_L D(x_L)}{\sqrt{1+\left(\kappa \delta R_\alpha^R/R_{R\uparrow}^L\right)^2 \left(1-x_L^2\right)}}, \qquad (7)$$

$$\lambda_L = \frac{1}{\kappa R_\alpha^L} \arctan(\kappa R_\alpha^L), \qquad (8)$$

$$\lambda_R = \frac{1}{\kappa R_\alpha^R/R_{R\uparrow}^L} \arctan(\kappa R_\alpha^R/R_{R\uparrow}^L), \qquad (9)$$

$$\tilde{\lambda}_1 = \int_{x_c}^{1} dx_L \frac{D(x_L)}{\sqrt{1+\left(\kappa R_\alpha^L\right)^2 \left(1-x_L^2\right)}}, \qquad (10)$$

$$\tilde{\lambda}_2 = \int_{x_c}^{1} dx_L \frac{\delta x_L D(x_L)}{\sqrt{x_L^2 \pm x_{cr}^2}\sqrt{1+\left(\kappa \delta R_\alpha^R/R_{R\uparrow}^L\right)^2 \left(1-x_L^2\right)}}, \qquad (11)$$

$$\tilde{\lambda}_3 = \int_{x_c}^{1} dx_L \frac{\delta x_L D(x_L)}{\sqrt{x_L^2 \pm x_{cr}^2}\sqrt{1+\left(\kappa R_\alpha^L\right)^2 \left(1-x_L^2\right)}}, \qquad (12)$$

$$\tilde{\lambda}_4 = \int_{x_c}^{1} dx_L \frac{D(x_L)}{\sqrt{1 + \left(\kappa \delta R_{\alpha}^{R}/R_{R\uparrow}^{L}\right)^2 \left(1 - x_L^2\right)}}, \tag{13}$$

where $R_{\alpha}^{L} = l_{L\alpha}/l_{L\uparrow}$, $R_{R\alpha}^{L} = l_{L\alpha}/l_{R\alpha}$, $R_{\alpha}^{R} = l_{R\alpha}/l_{R\uparrow} = R_{\alpha}^{L} R_{R\uparrow}^{L}/R_{R\alpha}^{L}$, $\alpha = (\downarrow, \uparrow)$, and the momentum conservation law $p_F^L \sin\theta_L = p_F^R \sin\theta_R$ has been used to bring the integration variables to the left-hand side incidence angle $\theta_L$. If $p_F^L < p_F^R$ then $x_c = 0$, $x_{cr} = \sqrt{(1-\delta^2)/\delta^2}$, where $\delta = p_F^L/p_F^R$, and the upper sign in the square roots of Eqs. (11) and (12) has to be used. When $p_F^L > p_F^R$, $x_c = x_{cr}$, $x_{cr} = \sqrt{1-(\delta)^{-2}}$, and the lower sign in the above mentioned square roots has to be used. Other notations are explained as follows: $p_F^i$ and $l_i$ are the Fermi momentum and the mean-free path of the $i$-th side ferromagnet, respectively; $a$ is a radius of the nanocontact, $J_1(\kappa a/l_{L\uparrow}) = J_1(ka)$ is the Bessel function; $\langle ... \rangle_{\theta_i}$ is the solid-angle averaging, and $x_i = \cos\theta_i$ is a cosine of the incidence angle on the $i$-th side ferromagnet, respectively.

The set of parameters, $R_{\alpha}^{L}$, $R_{\alpha}^{R}$, $R_{R\alpha}^{L}$, which determine relationships between the spin-dependent MFP, needs to be explained in more detail. The quantities $R_{\alpha}^{L}$ and $R_{\alpha}^{R}$ are formal spin asymmetries of the bulk MFP for the left- and right- hand sides, respectively. If the spin index $\alpha$ coinsides with a spin index of a mean-free path in the system, used for normalization ($l_{L\uparrow}$ in the actual case), then $R_{\uparrow}^{L} = R_{\uparrow}^{R} = 1$. For the opposite spin projection, $R_{\downarrow}^{L} = l_{L\downarrow}/l_{L\uparrow}$, $R_{\downarrow}^{R} = l_{R\downarrow}/l_{R\uparrow}$ are the material parameters which can be related to the bulk spin asymmetry coefficient [27],

$$\beta = \frac{\rho_\downarrow - \rho_\uparrow}{\rho_\uparrow + \rho_\downarrow} = \frac{1 - \sigma_\downarrow/\sigma_\uparrow}{1 + \sigma_\downarrow/\sigma_\uparrow}, \tag{14}$$

as follows:

$$\beta_{L(R)} = \frac{1 - \delta_{L(R)}^2 R_{\downarrow}^{L(R)}}{1 + \delta_{L(R)}^2 R_{\downarrow}^{L(R)}}, \text{ because } \sigma_{\alpha} \propto p_{F\alpha}^2 l_{\alpha}. \tag{15}$$

In equation (15), $\delta_{L(R)} = p_{F\downarrow}^{L(R)}/p_{F\uparrow}^{L(R)}$ characterizes the spin polarization of the conduction band of the left- (right-) hand side ferromagnetic metal.

In addition, the conductance of the contact depends on the ratio of the left- to the right- hand side mean-free paths, $R_{R\alpha}^{L}$. Then, provided that one of the MFP is known, for example, $l_{L\uparrow}$, the other three can be retrieved with the use of the three parameters: $R_{\downarrow}^{L}, R_{\downarrow}^{R}, R_{R\downarrow}^{L}$. A somewhat complicated notations is the cost for the universality of the set of equations (1)- (13), describing four spin channel conductances which appear for the parallel (P) and antiparallel (AP) alignment of magnetizations (see below).

To account for a finite length of the nanocontact, we put a linear-profile domain wall in the constriction of the nanocontact [20-23]. The angular- and spin-dependent quantum – mechanical coefficient of transmission $D$ through the linear domain wall reads:

$$D^{SL}(x_L, L) = \frac{4 p_M p_m t^2(L) \pi^{-2}}{\left(p_M \beta - p_m \gamma\right)^2 + \left(p_M p_m \alpha + \chi\right)^2}, \tag{16}$$

where

$$\alpha = \text{Ai}(q_1 L) \text{Bi}(q_2 L) - \text{Bi}(q_1 L) \text{Ai}(q_2 L),$$

$$\beta = t(L)\{\text{Ai}(q_1 L)\text{Bi}'(q_2 L) - \text{Bi}(q_1 L)\text{Ai}'(q_2 L)\},$$
$$\gamma = t(L)\{\text{Ai}'(q_1 L)\text{Bi}(q_2 L) - \text{Bi}'(q_1 L)\text{Ai}(q_2 L)\}, \quad (17)$$
$$\chi = t^2(L)\{\text{Ai}'(q_1 L)\text{Bi}'(q_2 L) - \text{Bi}'(q_1 L)\text{Ai}'(q_2 L)\},$$

and $t(L) = [2mE_{ex}/L]^{1/3}$, $E_{ex} = (p_{FM}^2 - p_{Fm}^2)/2m$, $q_1 = -p_{FM}^2 t(L)/2mE_{ex}$, $q_2 = -p_{Fm}^2 t(L)/2mE_{ex}$, where $L$ is a width of DW; $\text{Ai}(z)$, $\text{Bi}(z)$, $\text{Ai}'(z)$, and $\text{Bi}'(z)$ are the Airy functions and their derivatives; $p_m = p_{Fm}\cos(\theta_m)$ and $p_M = p_{FM}\cos(\theta_M)$ are the normal components of the wave vector of minority and majority subband, respectively. Note here that $p_m$ is used for a subband with a smaller Fermi momentum, while $p_M$ for a subband with a larger Fermi momentum whatever the spin projection of the subband, or the side of the contact - left- or right-hand, is. The quantum-mechanics textbook expression for a coefficient of the transmission through the step-like DW (band-offset model), $D^{step}(x_L) = 4 p_M p_m / (p_M + p_m)^2$, can be retrieved from Eq. (16) in the limit of $L \to 0$. Here, we omit the spin index to simplify appearance of the above formulas.

The magnetoresistance of a magnetic nanocontact can be calculated as follows:

$$MR = \frac{\sigma^P - \sigma^{AP}}{\sigma^{AP}}, \quad (18)$$

where $\sigma^{P(AP)} = \sigma_{LR\uparrow}^{P(AP)} + \sigma_{LR\downarrow}^{P(AP)}$ is conductance for a parallel, $P$ (antiparallel, $AP$), mutual alignment of magnetization in the contacting ferromagnets. $\sigma_{LR\uparrow}^{P(AP)}$ ($\sigma_{LR\downarrow}^{P(AP)}$) is conductance of the spin-up (spin-down) conduction electron spin-channel, equation (1), at the parallel (antiparallel) alignment. Then, MR is positive if the physical effect itself is negative (resistance drops when magnetic field is applied). Now, the dependence of MR on the spin-dependent bulk electron MFP can be investigated for physically distinguished combinations of ferromagnetic metals.

In a simple parabolic band structure we use here, the heterocontacts are sorted out by the mutual positions of bottoms of their conduction bands at the parallel alignment of magnetizations. The first case of the same ferromagnetic material in a contact seems to be trivial, however, one side can differ substantially in electron scattering compared with the other. In fact, $3d$ and $4d$ solutes in iron [24,25] may result in the dramatic increase of the spin asymmetry of the conduction-electron scattering, whereas the Fermi momenta of the spin subbands remain practically unchanged because of the low impurity concentration. As a result, in our calculations we can keep band structure parameters (Fermi momenta and spin polarizations) of the contacting ferromagnets unchanged, but vary the spin-dependent MFP of the sides independently in a wide range.

Figure 1 displays the calculated dependences of MR on the contact radius for a set of MFP spin asymmetries. The contact length $L$ is chosen equal to 10 Å (1 nm). The general conclusion is that MR rapidly drops as the contact size increases beyond the mean-free path for the either spin channel of conductance. It is an indication that to enhance MR one has to strive towards the ballistic regime of conductance in vicinity of the contact. On the other hand, our calculations give a hint that for certain combinations of the mean-free path asymmetries, Figs. 1b, 1e, and especially Fig. 1c, the reduction of MR with increasing the contact cross-section size is not so drastic, and the requirement of the nanometric contact size to reach high MR values is not so strict.

For heterocontacts, three physically distinct combinations are considered (see insets in Figs. 2-4), and MR for every combination is calculated against the contact radius for a set of MFP values. Figure 2 shows the case when one ferromagnet has an essentially higher conduction electron density compared with the other. The absolute MR values are not so exciting, especially from the point of view of magnetic field sensor applications, however, MR Eq. (18), can be not only positive but also negative, depending on the combination of the spin asymmetries of the contacting ferromagnets (see Fig. 2b, 2c and 2e). Moreover, Figs. 2a, 2b, 2d and 2e show non-monotonous behavior $MR(a)$ with a minimum between $a$ and $3a$, which we attribute to the mixed regime of conductance in the vicinity of the contact: ballistic for one spin channel ($a/l_{L\downarrow} \ll 1$), and diffusive for the other ($a/l_{R\uparrow} \gg 1$).

Figure 3 shows the case when one ferromagnet has a bit higher conduction electron density than the other. The absolute MR values ~ 130% are much higher in the ballistic limit ($a/l_{L\uparrow} \ll 1$) than in the previous case. Again, Figs. 3d and 3e show the non-monotonous behavior of MR as the function of the contact radius. Figures 3b, 3c and 3e suggest combinations of parameters, for which MR keeps high values with increasing the contact cross-section.

Figure 4 (a to d) shows the case of the contact when ferromagnets have similar conduction electron densities but different spin polarizations of the conduction band. Although MR magnitudes are moderate, Figs. 4b and 4c suggest a weak dependence of MR on the contact radius at certain combinations of the MFP spin asymmetries (see the figure caption).

We do investigate MR curves not only with set of $\left(l_{L\uparrow}, R_{\downarrow}^{R}, R_{R\downarrow}^{L}, R_{\downarrow}^{R}\right)$ (Fig.1-Fig.4), but also from another one set $\left(l_{L\uparrow}, R_{\downarrow}^{R}, R_{R\uparrow}^{L}, R_{\downarrow}^{R}\right)$ (see Fig.5-Fig.7). These two sets uniquely define the set of MFP values.

Concluding the chapter, we note that in contrast to the case of large-area thin film contacts [26], where MR does not depend on mean-free paths, in the case of a nanocontact we have strong dependence on the spin asymmetry as well as on the absolute value of the MFP. The different behavior is explained by the shrinkage of a current to a nanocontact size of the same scale as a mean-free path. Then, if the spin asymmetries are large, one of the spin channels of conduction can be ballistic, but the other one, or the matching spin channel of the opposite lead of the junction, can be in the diffusive regime either. Upon the change of the contact size the conduction regimes in the two spin channels change gradually, that makes MR dependent on the mean-free paths. In the case of two flat layers, the current flows homogeneously through the interface of a formally infinite lateral dimension. Then, there is no a structural dimension comparable with the mean-free paths that makes the short-scale MFP effects ineffective.

## 3. Discussion and Conclusions

To the best of our knowledge there are two reports on MR of ferromagnetic heterocontacts with metallic conductance between Mumetal and Ni [10,17]. Mumetal ($Ni_{77}Fe_{14}Cu_{5}Mo_{4}$) is close to Permalloy ($Ni_{80}Fe_{20}$) in its composition, hence, it has a very short mean-free path in the spin-down conduction channel [15,16,28,29]. Therefore, we refer Mumetal-Ni heterocontact as corresponding to the third case (Fig. 4). For the numerical calculations of MR we used MFP of conduction electrons in permalloy [15,29,30] as guess values for Mumetal, and MFP in nickel estimated from the data of Ref. [30]. The larger spin-dependent Fermi momentum in Ni has been assigned to the spin-down subband which has higher density of states according to the spin-polarized density-of-state calculations (see, for example, [31,32]). Results of the calculations are given in Fig. 4e. For the ballistic regime of conductance ($a/l_{L\downarrow} \to 0$) MR is close to 100% ($L$ = 0.5 nm) which agrees satisfactorily with experimental MR values = 78-132%, Fig. 2 in [10], at lowest conductances for the P-alignment of magnetizations.

To summarize, in this paper we investigated mean-free path effects on MR of ferromagnetic nanocontacts. Ferromagnetic heterocontacts mean that the contacting ferromagnets have different parameters of their conduction bands. In our calculations, we fix spin-polarization parameters of the ferromagnetic metal at the left bank of the contact $p_{F\downarrow}^{L} = 0.4$ Å$^{-1}$, $p_{F\uparrow}^{L} = 1.0$ Å$^{-1}$, $\delta_{L} = 0.4$ which are close for iron, and vary the conduction band properties of the second ferromagnetic metal. In most cases the MR monotonously decreases as the contact cross-section increases. For some cases with a big difference in spin subband mean-free paths, the calculated MR shows non-monotonous behavior in the region where the diameter of the contact becomes comparable with the mean-free path of electrons. We attribute this effect to the gradual change of conduction regimes in vicinity of the nanocontact upon changing the contact cross-section size. As a result, at certain combinations of the mean-free paths in heterocontacts, the MR can be almost constant or even increase a bit as the contact size increases. The latter findings open a way to search for proper solutes for ferromagnetic materials, which provide conditions to realize high MR at technologically available cross-sections of the nanocontacts. The trial calculations of the magnetoresistance with material parameters close to that for the Mumetal-Ni heterocontacts agree satisfactorily with the available experimental data.

## 4. Acknowledgments

The work was supported by EC through the project NMP4-CT-2003-505282, and by the Russian Ministry of Science and Education.

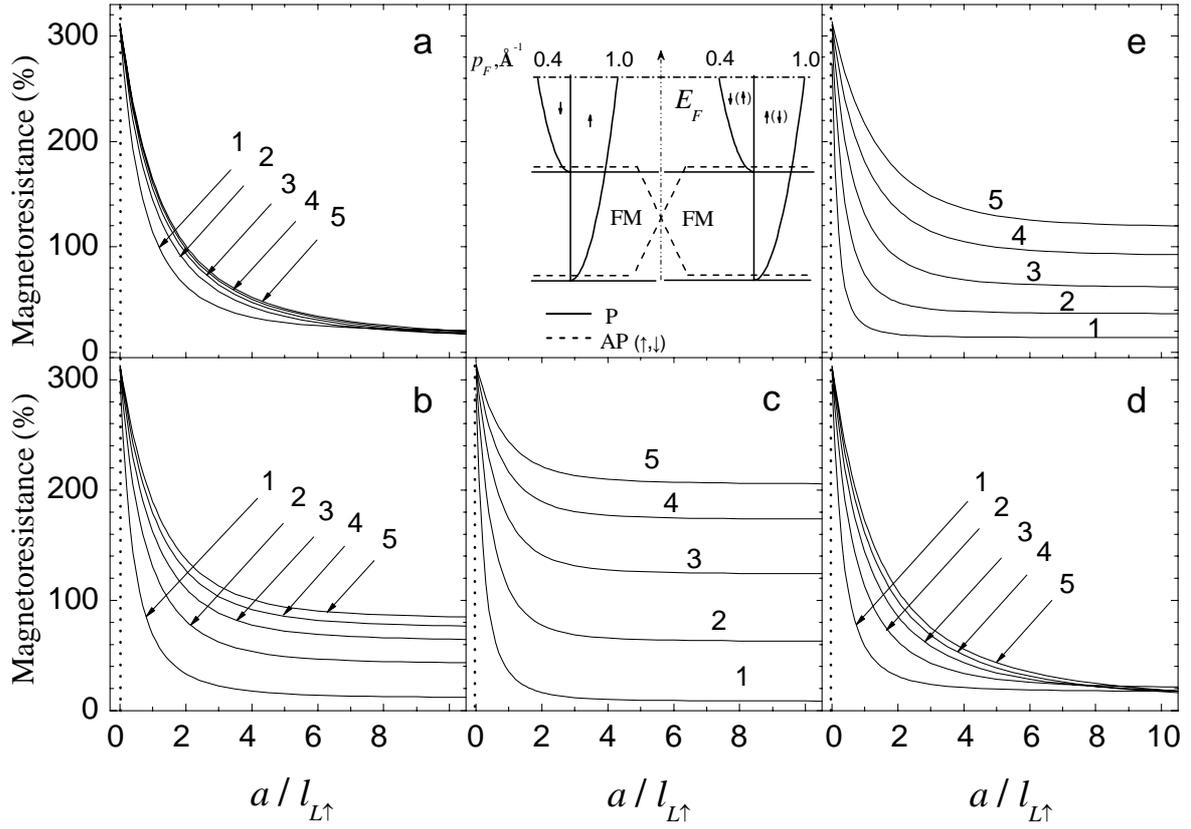

**Fig.1.**

Dependence of MR on the contact radius for the case of homocontact ($\delta_L = \delta_R = 0.4$).

The mean-free paths ratio is:

a) $R^L_\downarrow = 5.0$, $R^L_{R\downarrow} = 1.0$;   c) $R^L_\downarrow = 1.0$, $R^L_{R\downarrow} = 1.0$;   e) $R^L_\downarrow = 2.0$, $R^L_{R\downarrow} = 3.0$

b) $R^L_\downarrow = 2.0$, $R^L_{R\downarrow} = 1.0$;   d) $R^L_\downarrow = 5.0$, $R^L_{R\downarrow} = 3.0$;

$R^R_\downarrow = 5.0, 2.0, 1.0, 0.5, 0.2$ in accordance with the labels set from 1 to 5. $L = 1.0$ nm.

All MFP values are determine from $\left(l_{L\uparrow}, R^R_\downarrow, R^L_{R\downarrow}, R^R_\downarrow\right)$, where $l_{L\uparrow} = 3$ nm.

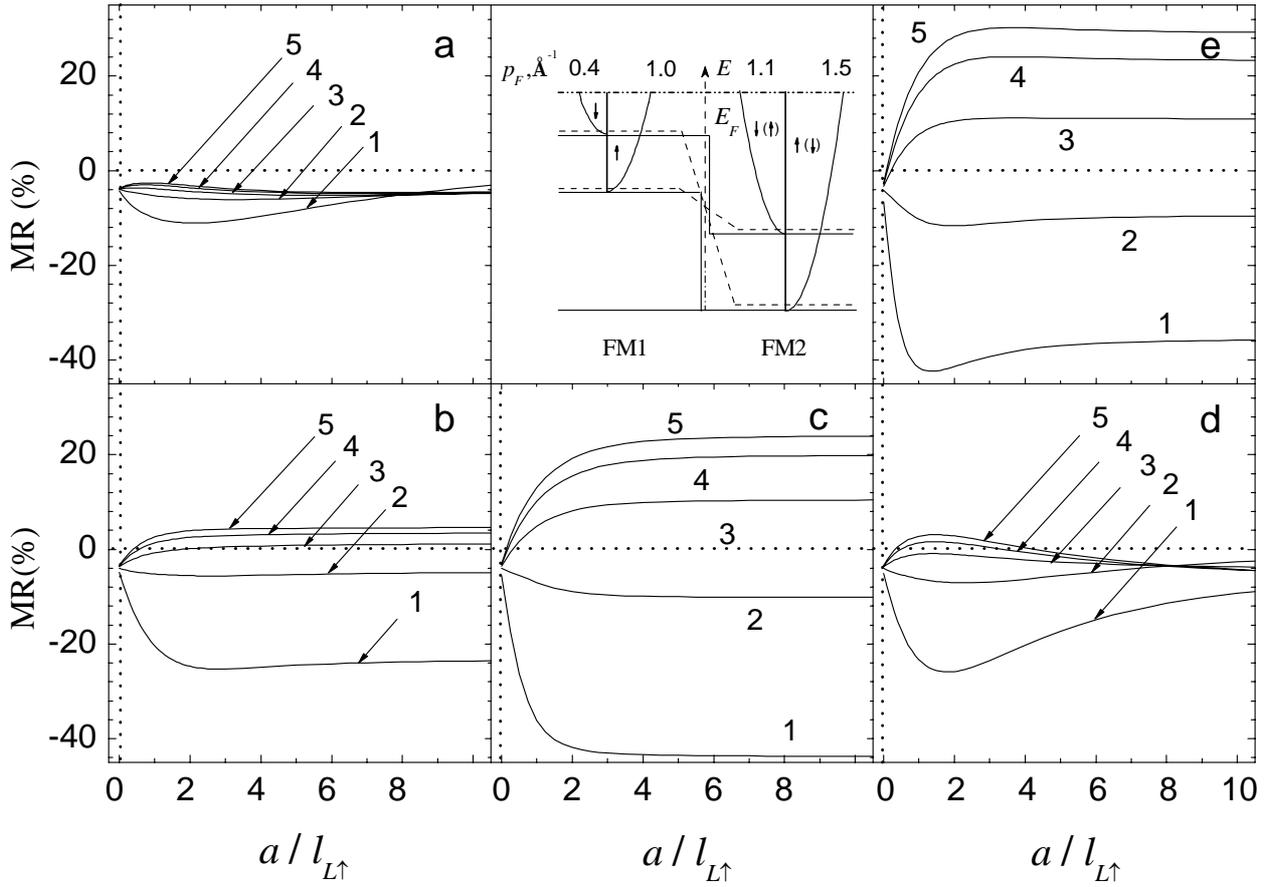

**Fig. 2.**

Dependence of MR on the contact radius for the case of heterocontact with $p_{F\downarrow}^{L} < p_{F\uparrow}^{L} < p_{F\downarrow}^{R} < p_{F\uparrow}^{R}$ ($\delta_L = 0.4$, $\delta_R \simeq 0.73$). The layout and choice of the mean-free paths are the same as in Fig.1.

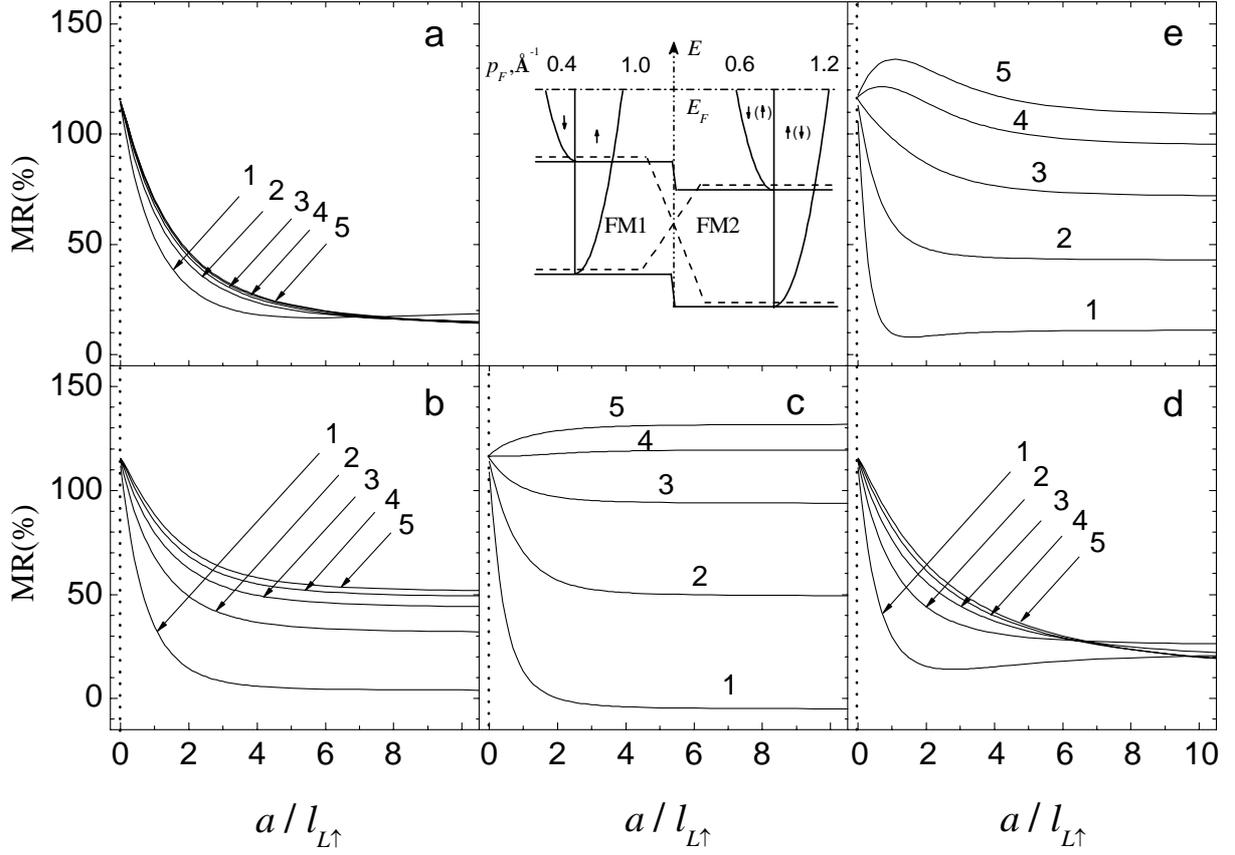

**Fig. 3.**

Dependence of MR on the contact radius for the case of heterocontact with $p_{F\downarrow}^L < p_{F\downarrow}^R < p_{F\uparrow}^L < p_{F\uparrow}^R$ ($\delta_L = 0.4$, $\delta_R = 0.5$). The layout and choice of the mean-free paths are the same as in Fig.1.

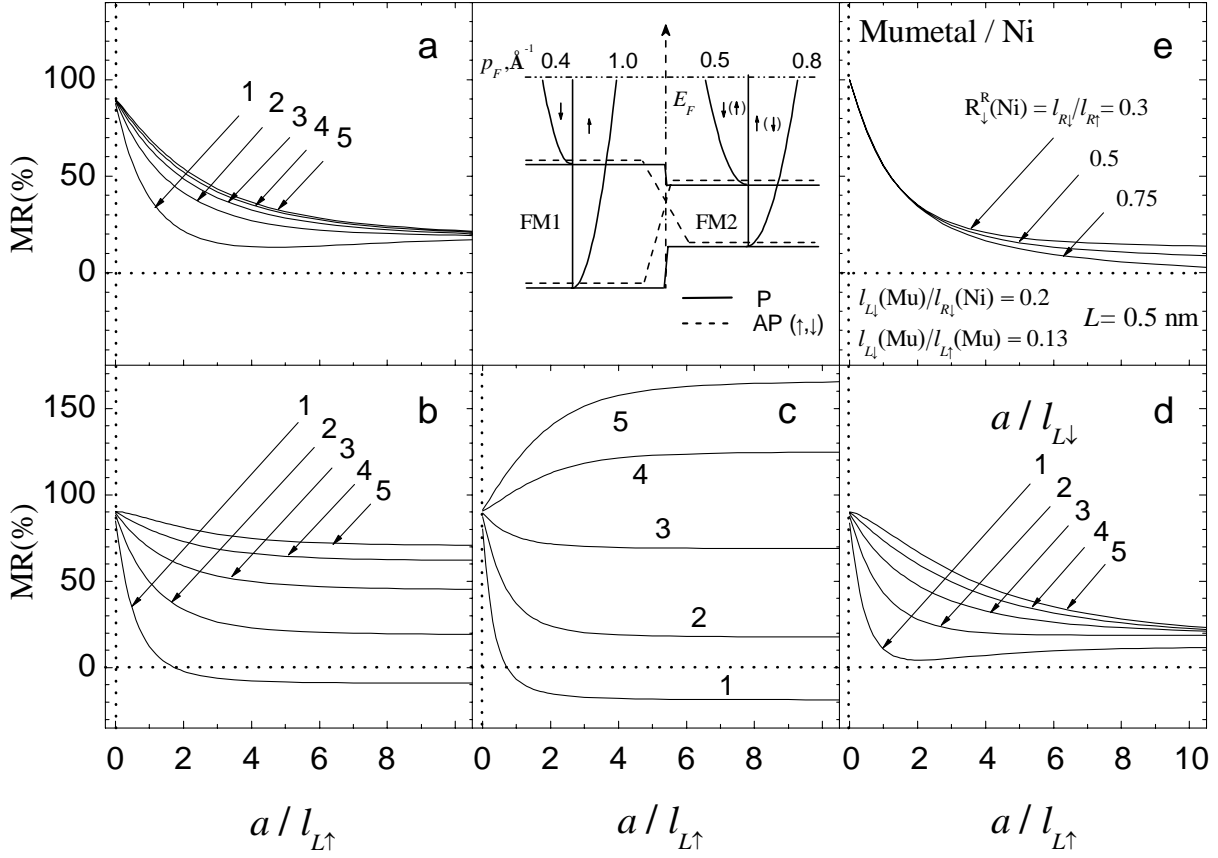

**Fig. 4.**

Dependence of MR on the contact radius for the case of heterocontact with $p_{F\downarrow}^L < p_{F\downarrow}^R < p_{F\uparrow}^R < p_{F\uparrow}^L$ ($\delta_L = 0.4$, $\delta_R \simeq 0.63$). The choice of the mean-free paths are the same as in Fig. 1 except for the plot **e**).

The plot **e**) is the dependence of MR on the contact radius for Mumetal-Ni heterocontact ($p_{F\uparrow}^L(\mathrm{Mu}) = 0.61\,\text{Å}^{-1}$, $p_{F\downarrow}^L(\mathrm{Mu}) = 1.1\,\text{Å}^{-1}$, $p_{F\uparrow}^R(\mathrm{Ni}) = 0.65\,\text{Å}^{-1}$, $p_{F\downarrow}^R(\mathrm{Ni}) = 1.08\,\text{Å}^{-1}$) $l_{L\downarrow}(\mathrm{Mu}) = 0.6$ nm and the parameters of mean-free path ratios are as follows: $R_\downarrow^L = 0.13$ [15,29], $R_{R\downarrow}^L = 0.2$ [15,29,30], $R_\downarrow^R = 0.3, 0.5$ and $0.75$ in accordance with the possible range for Ni [30].

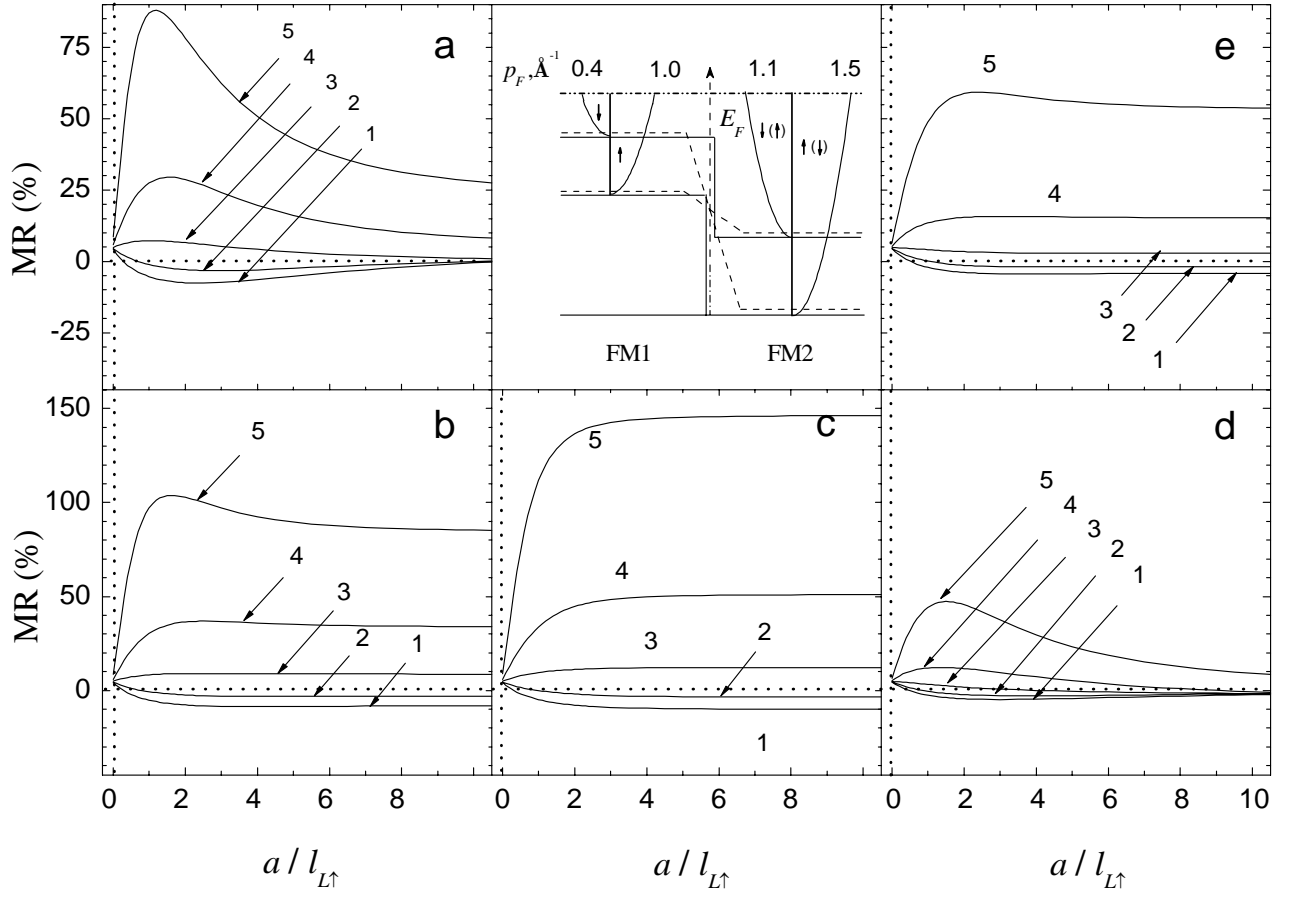

**Fig. 5.**

Dependence of MR on the contact radius for the case of heterocontact with $p_{FL\downarrow(\uparrow)} < p_{FR\uparrow(\downarrow)}$ ($\delta_L = 0.4$, $\delta_R \simeq 0.73$).

The mean-free paths ratio is:

a) $R^L_\downarrow = 5.0$, $R^L_{R\uparrow} = 1.0$;   c) $R^L_\downarrow = 1.0$, $R^L_{R\uparrow} = 1.0$;   e) $R^L_\downarrow = 2.0$, $R^L_{R\uparrow} = 0.5$

b) $R^L_\downarrow = 2.0$, $R^L_{R\uparrow} = 1.0$;   d) $R^L_\downarrow = 5.0$, $R^L_{R\uparrow} = 0.5$;

$R^R_\downarrow = 5.0, 2.0, 1.0, 0.5, 0.2$ in accordance with the labels set from 1 to 5.   $L = 1.0$ nm.

All MFP values are determine from $\left(l_{L\uparrow}, R^R_\downarrow, R^L_{R\uparrow}, R^R_\downarrow\right)$, where $l_{L\uparrow} = 3$ nm.

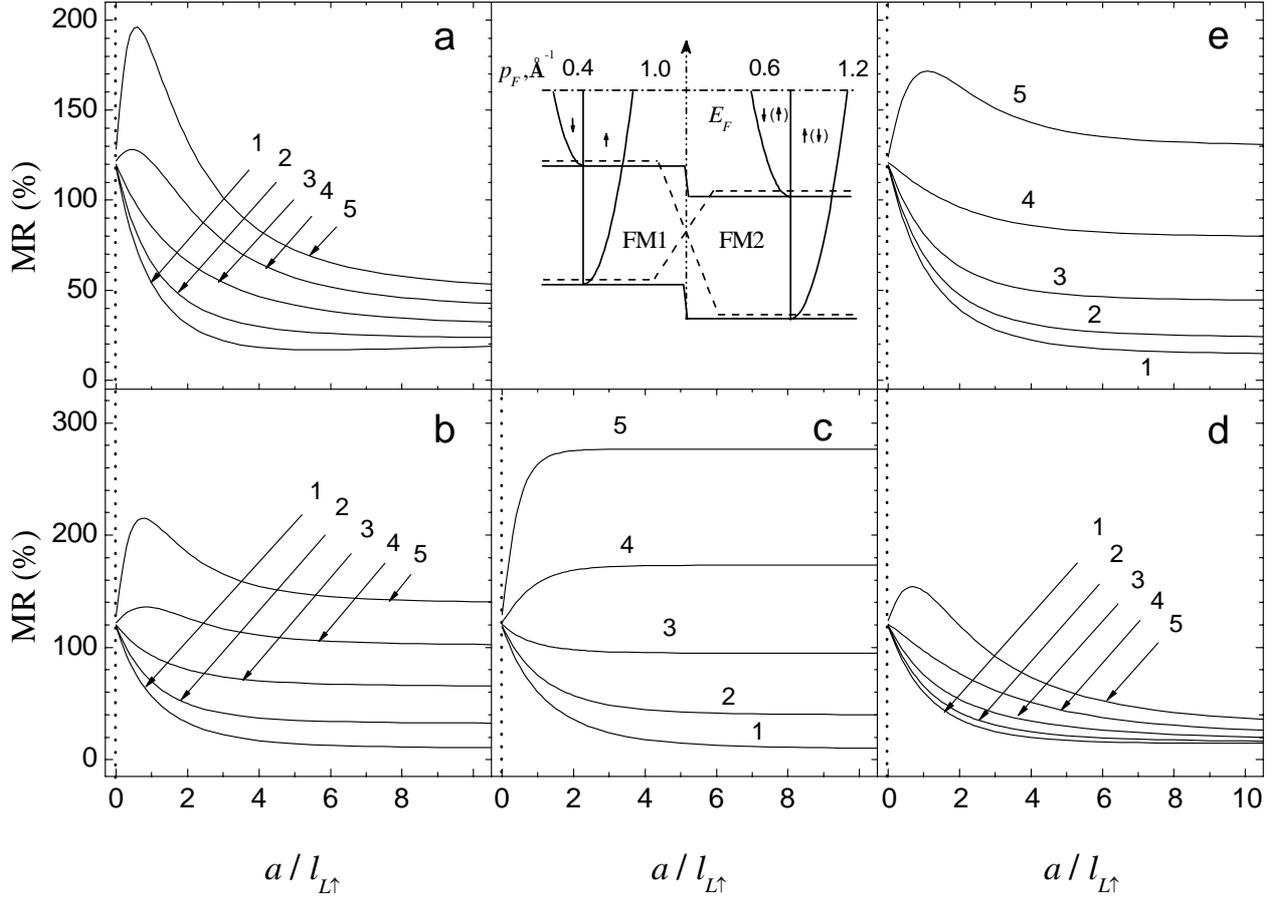

**Fig. 6.**

Dependence of MR on the contact radius for the case of heterocontact with $p_{F\downarrow}^L < p_{F\downarrow}^R < p_{F\uparrow}^L < p_{F\uparrow}^R$ ($\delta_L = 0.4$, $\delta_R = 0.5$). The layout and choice of the mean-free paths are the same as in Fig.5.

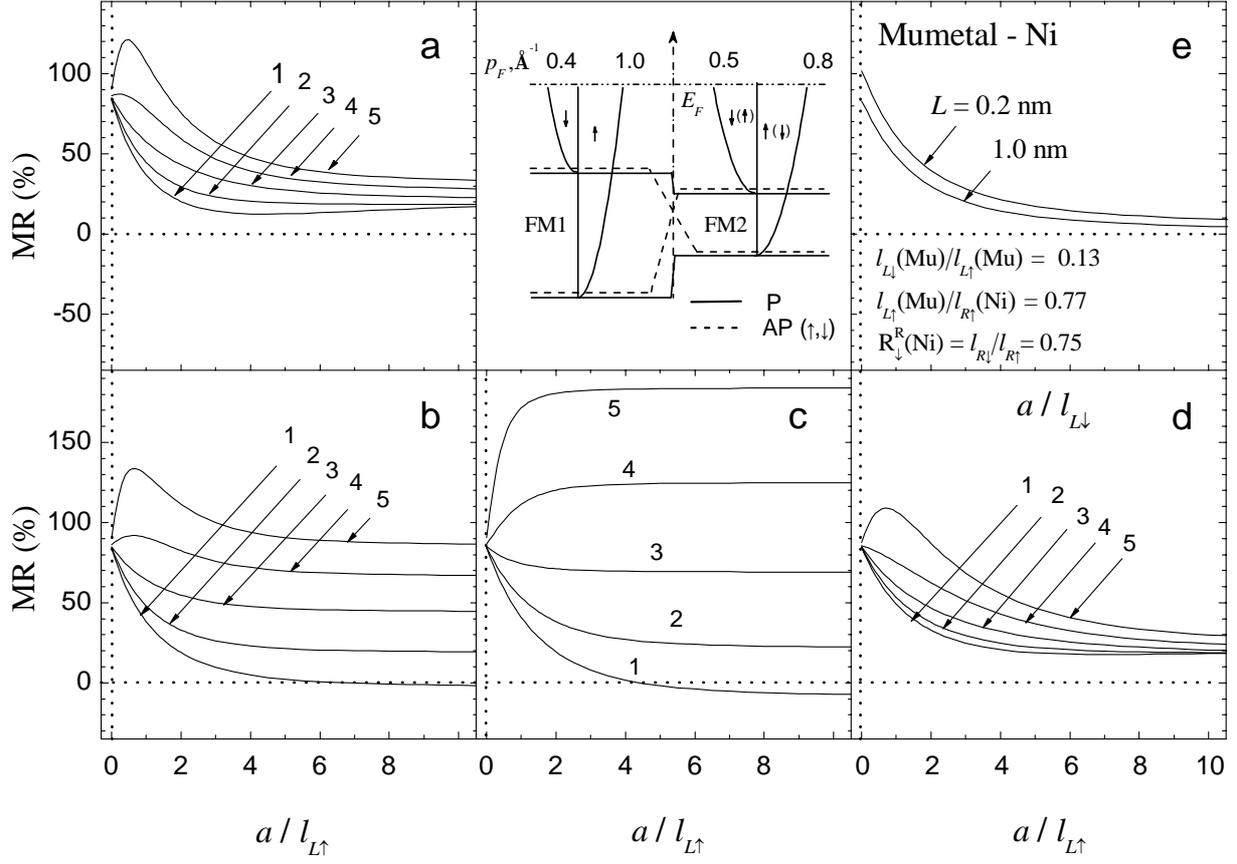

**Fig. 7.**

Dependence of MR on the contact radius for the case of heterocontact with $p_{F\downarrow}^L < p_{F\downarrow}^R < p_{F\uparrow}^R < p_{F\uparrow}^L$ ($\delta_L = 0.4$, $\delta_R \simeq 0.63$).

The layout and choices of the mean-free paths are the same as in Fig. 5 except for the plot **e**).